\documentclass{egpubl}
\usepackage{eurovis2026}

\CGFccbyncnd

\usepackage[T1]{fontenc}
\usepackage{dfadobe}

\usepackage{cite}  
\usepackage{multirow}
\usepackage{array}
\usepackage{booktabs}

\BibtexOrBiblatex
\electronicVersion
\PrintedOrElectronic
\ifpdf \usepackage[pdftex]{graphicx} \pdfcompresslevel=9
\else \usepackage[dvips]{graphicx} \fi

\newcommand{\npaper}{318}
\newcommand{\npublication}{333}
\newcommand{\nconceptualization}{160}
\newcommand{\napplication}{159}

\title{What Do We Mean When We Talk about Data Storytelling?}

\author[L. Yang \& Z. Wang \& S. Carpendale \& X. Lan]
{
{\parbox{\textwidth}{\centering L. Yang$^{1}$\orcid{0000-0003-4527-4905}, Z. Wang$^{2}$\orcid{0000-0002-1061-604X}, S. Carpendale$^{2}$\orcid{0000-0002-5127-9780}, X. Lan\thanks{Corresponding author}$^{3}$\orcid{0000-0001-7331-2433}
}}
        \\
{\parbox{\textwidth}{\centering $^1$ Inria, CNRS\\
         $^2$ School of Computing Science, Simon Fraser University\\
         $^3$ School of Journalism, Fudan University
       }
}
}

\begin{document}


\maketitle
\begin{abstract}
Data storytelling has seen rapid growth through a proliferation of examples, as well as theoretical and technical advancements contributed across multiple disciplines. In this paper, we present a comprehensive survey of data storytelling research from 2010 to 2025. By analyzing the conceptualizations of data storytelling collected from related publications, we reveal the field's perspectives on the \emph{What, How, Why}, and \emph{Who} of data storytelling. We further investigated the operationalization of data stories. We identified 12 data story forms that provide concrete examples of how data stories have been presented. We derived a set of spectrum-based dimensions that capture important properties of data stories. Along each spectrum, applicable forms and design alternatives were discussed to analyze how they shape data storytelling experiences, along with data storytelling design trade-offs. Additionally, we examine how traditional narrative elements, like plot and character, have been adapted in data stories to support the operationalization of a data storytelling narratological perspective. Finally, we concluded the survey with a synthesis of our major findings and implications for future research.



\end{abstract}  

\section{Introduction}
For the visualization community, storytelling with data has become an increasingly prominent topic. In recent years, a growing body of work has examined its theories, design spaces, and enabling techniques~\cite{tong2018storytelling,m7zhao2023stories,chen2023does,shen2025does}. In practice, data storytelling is also widely adopted by journalists, designers, and visualization practitioners. Communities such as DataJournalism.com~\cite{DJ} and the European Data Journalism Network~\cite{EJ} curate and share methods and examples, while books such as \textit{Data-Driven Storytelling}~\cite{b13riche2018data} discusses the expanding possibilities of including data in stories and \textit{Storytelling with Data}~\cite{knaflic2015storytelling} offers practical advice on refining charts for effective communication, especially in business contexts.

However, the rapid proliferation of work around data storytelling has also surfaced ambiguity about what scholars and practitioners mean when they refer to Data Stories. Multiple labels, such as \emph{data story}, \emph{data(-driven) storytelling}, and \emph{narrative visualization}, are often used interchangeably. Some accounts define data stories primarily through narrative structure, emphasizing a coherent progression with a beginning, middle, and end~\cite{a17yang2021design}. Others place greater emphasis on agency and interactive exploration, where meaning can be constructed through engagement with the visualization rather than through a fully pre-authored sequence~\cite{a1segel2010narrative}. Core concepts are also used inconsistently across venues and communities. For example, \emph{character} can refer to literal people in the underlying data, or it can be used metaphorically to describe data entities that act as protagonists within a narrative framing~\cite{a472dasu2023character}. These differences are not merely terminological. They shape how stories are designed, how systems are evaluated, and how findings are communicated.

At the same time, data storytelling practices continue to diversify in both medium and setting. Beyond combinations of text, images, and video, recent work has expanded toward immersive and situated experiences, including Virtual Reality (VR), Augmented Reality (AR), and Extended Reality (XR), as well as experiments with live performance and other artistic forms~\cite{m7zhao2023stories}. This breadth is promising, but it also makes it difficult to maintain a coherent picture of the field. A survey can explore ways to map how data storytelling is conceptualized, to synthesize connections across domains and communities, and to reflect on future possibilities without losing the larger landscape.

To fill this gap, our survey codes how data storytelling is conceptualized in research publications from January 2010 to November 2025 (N = \npublication). Our corpus spans more than ten application domains and over seven subject areas, reflecting the breadth of data storytelling research (\autoref{fig:paper_overview}). Our analysis identifies five recurring perspectives (i.e., \textit{technique-and-goal}, \textit{visualization-centric}, \textit{technique-centric}, \textit{goal-centric}, and \textit{interdisciplinary}). Across these perspectives, we observe multifaceted understandings in data storytelling based on What is told in data storytelling? (\textbf{What}), What are the techniques used in data storytelling? (\textbf{How}), What effects should data storytelling achieve? (\textbf{Why}), and Who is the audience of data storytelling? (\textbf{Who}).

To better understand how research operationalizes data stories, we examine the literature from two angles. First, we catalogue the \textit{forms} used to present data stories. We identify 12 forms (e.g., comic strip, video, article, etc.), which provide examples of what data stories can be presented. At the same time, we find that form-based typologies often suffer from ambiguity, overlap, and limited flexibility. We therefore complement forms with four spectrum-based dimensions that describe properties shaping the audience experience: \textit{spatiality} (1D--3D), \textit{ordering} (how the experience is sequenced and navigated, from ordered progression to free-form exploration), \textit{sensory modalities} (uni- to multisensory), and \textit{audience co-authorship} (low--high). For each dimension, we discuss trade-offs and illustrate design choices with examples.

Second, because \emph{narrative} is frequently invoked as a defining property of data storytelling, we analyze how three core narrative elements---event, plot, and character---manifest in data stories. We show that these elements are often reinterpreted in ways that differ from classic narratology, and we use these observations to reflect on the nature of data storytelling from a narratological perspective. Finally, we synthesize insights across our analyses and outline our discoveries of implications and opportunities for future work.

In summary, our contributions are: (1) a fresh synthesis of how data storytelling has been conceptualized in research publications across visualization and related disciplines; (2) a narratological analysis of how event, plot, and character are operationalized in data stories; (3) a taxonomy of 12 data story forms and four experience-oriented spectra (spatiality, sensory modalities, ordering, and audience co-authorship) that support cross-genre comparison and design reasoning; and (4) implications and discussions for future research and practice.

\section{Background}
In this section, we first compare our survey with existing surveys in data storytelling.
We then explain how our approach of reviewing conceptualizations is inspired by related research in conceptualization in the visualization community.

\subsection{Literature Review of Data Storytelling}
\label{sec:literature}
The first known comprehensive survey is by Tong et al. \cite{tong2018storytelling}. They classified and discussed literature based on \textit{Who}, \textit{How}, and \textit{Why}, and the path the viewer takes through the visualization (sub-categories: \textit{Linear}, \textit{audience-directed path}, \textit{Parallel}, and \textit{Random access or other}). 
This offered a preliminary understanding of data stories. 
However, their survey scope was limited to the fields of engineering and computer science, including papers in IEEE and ACM Digital libraries.
New forms and techniques of data stories have been emerging across different domains in recent years. 
For example, their \textit{How} dimension includes two sub-categories: \textit{narratives} and \textit{transitions}, 
This prompts us to reflect on what new concepts are proposed regarding the dimensions mentioned in this work across a broader timeline and domains. 

More literature surveys have been published in the past five years for different perspectives. Many surveys surround how authoring tools can support data storytelling creations~\cite{li2024we,chen2023does,he2024leveraging,ye2025data,garreton2025survey}. They focus on technical issues and interaction paradigms of authoring tools. For example, He et al.~\cite{he2024leveraging} investigate what foundation models can help different tasks (e.g., Insight Extraction and Authoring) in creating narrative visualizations. Li et al.~\cite{li2024we} summarized different roles human and AI can perform in a collaborative way to create data stories.
Some surveys focus on data storytelling in particular domains, including data journalism~\cite{fu2023more,freixa2021binomial,francesca2025different} and business analysis~\cite{quenum2025enhancing}. Other surveys focus on the evaluations of data stories~\cite{a348errey2023evaluating,airaldi2025best}. 

The most closely related survey is by Schröder et al.~\cite{Schroeder2023Telling}. They reviewed the theoretical and conceptual foundations of data-driven storytelling, summarizing frameworks of the data storytelling process and design guidelines. They classified the designs and techniques based on Chatman~\cite{chatman1978story}'s narrative model of narrative components, including \textit{characters}, \textit{settings}, \textit{structure}, \textit{elements}, and so forth. 
Given the scope and focus of their review, the paper places emphasis on a selected subset of frameworks. Building on this foundation, our study complements their synthesis by assembling a broader account of how data storytelling has been conceptualized across research communities. This more comprehensive view supports reflection on the space of data stories and helps identify opportunities to extend current practices.
In addition, we explicitly articulate the similarities and differences between the fields of narratology and data storytelling in how they interpret three core narrative elements, including event, plot, and character. This comparison suggests that existing narrative models may not fully accommodate the specific properties of data stories, motivating the development of a model tailored to this context.

\subsection{Conceptualization in the Visualization Community}
Visualization research is advancing at a remarkable pace, with its techniques, application domains, and target audiences becoming increasingly diverse.
This development introduces the emergence of new concepts, such as \emph{situated visualization}~\cite{bressa2021s}, \emph{affective visualization design}~\cite{lan2023affective}, and the refinement of established concepts such as \emph{insights}~\cite{burns2023we,chang2009defining,yi2008understanding}.
Reflections on conceptualization are of vital importance for prompting a consistent and coherent research landscape~\cite{bressa2021s}, understanding how existing research results are related~\cite{burns2023we}, and guiding the designs and evaluations of visualization techniques or tools~\cite{battle2023we, burns2023we,bressa2021s}.

Reviews of existing conceptualizations have primarily supported two broad types of research outcomes.
One type of research synthesized existing conceptualizations to propose a more inclusive conceptualization or theory.
For example, Dimara and Perin~\cite{dimara2019interaction} proposed an inclusive conceptualization of \emph{interaction} in data visualization by synthesizing conceptualizations from both the visualization and human-computer interaction communities.
Recently, Battle and Ottley~\cite{battle2023we} studied what is a data \emph{insight}. They synthesized theories and conceptualizations of data insights into a unified formalism, including the core building blocks of insights and a graph-based data model that structures these building blocks for specifying an insight.
Another type of research reflected on existing conceptualizations to derive current research practices, research scopes, and important research topics with implications for future research practices and agendas.
For example, Bressa et al.~\cite{bressa2021s} reviewed the conceptualizations of \emph{situated visualization} to understand the essence of \emph{situatedness}.
Through case analysis, they further discussed five perspectives on situatedness, including space, time, place, activity, and community.
Recently, Burns et al.~\cite{burns2023we} systematically reviewed the themes in conceptualizations of \emph{visualization novice} and the characteristics of participants in studies with visualization novices as the target users.
Based on the analysis, they suggested strategies for future work to clarify conceptualizations of novices, such as offering counter-examples. 

Our work is largely inspired by the second type of research. We derived themes from the current conceptualizations of data storytelling and analyzed how important elements (e.g., narrative structure and characters) of data storytelling are applied in data stories created or evaluated by related literature.
Inspired by Dimara et al.~\cite{dimara2019interaction} who also tackled highly cross-disciplinary issues, we did not limit our investigation to work published within the visualization community. Instead, we gathered and analyzed literature from various disciplines and fields. To the best of our knowledge, this is the first systematic review for characterizing the conceptualizations of data storytelling.

\begin{figure*}[h]
  \centering
  \includegraphics[width=\textwidth]{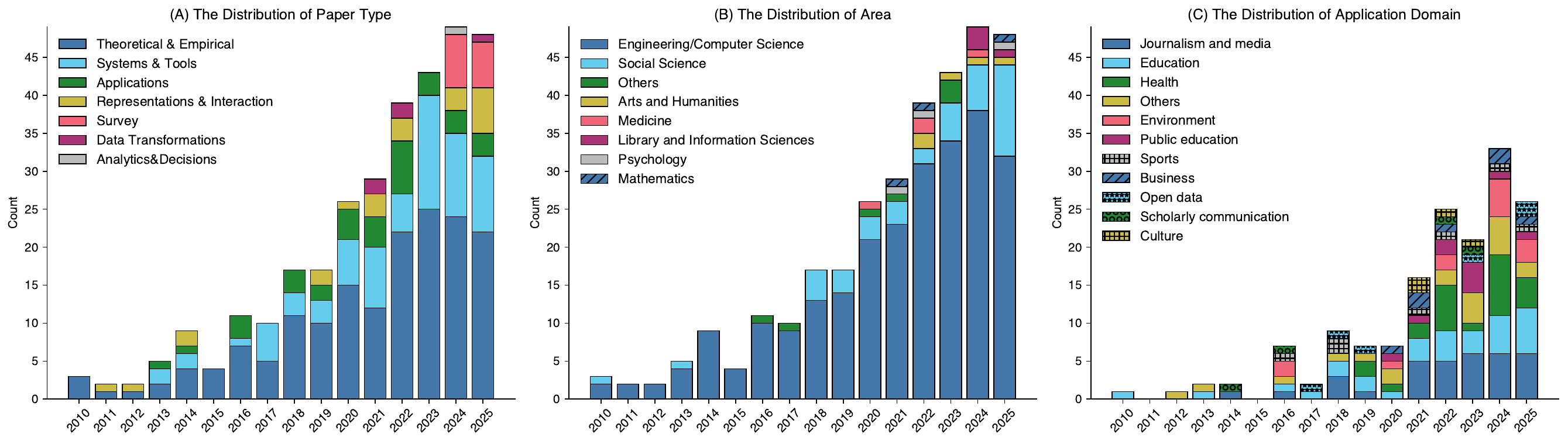}
  \caption{The distributions of the subject areas and application domains of the publications in our corpus from January 2010 to November 2025. Note that only \napplication~papers have explicitly claimed which domain(s) they anticipate data stories will be applied to.
}
  \label{fig:paper_overview}
\end{figure*}

\section{Methods Overview}
This section introduces how we constructed and analyzed a corpus of data storytelling, including the corpus collection process, the overview of the corpus, and how we collaboratively coded it. 

\subsection{Corpus Collection}
We collected academic publications published between January 2010 and November 2025 from mainstream academic databases, including IEEE, ACM, Scopus, Web of Science, and Taylor \& Francis. We selected 2010 as the start time, as Segel and Heer's~\cite{a1segel2010narrative} work in that year is a milestone when data storytelling is considered a standalone story form different from traditional storytelling. In 2010, there were a total of three pioneering works~\cite{a1segel2010narrative, pfannkuch2010telling,social2009}. Since then, publications in this field have increased remarkably. 
 
We applied keyword searching in the title and abstract. 
The set of keywords includes ``narrative visualization'', ``data(-driven) storytelling'', ``data story(-ies)'', ``storytelling with data'', ``data journalism'', ``data(-driven) articles'', and ``data(-driven) news''. 

In this survey, we treat data as recorded or collected evidence that a story draws on to support, explain, question, or test its claims. This includes quantitative and qualitative materials (e.g., texts, images, observations), but it does not include information that is only provided as background context or general narrative content. 
We filtered for publications that 1) consider data as the major subject in storytelling, 2) focus the research on data stories, such as authoring, evaluating, and analyzing data stories, and 3) are peer-reviewed and accessible. 
Excluded examples include publications that 1) use ``data-driven storytelling/story/article'' to represent AI-generated stories that are learned from datasets of stories instead of stories about data, 2) refer to ``narrative visualization'' as a type of data visualization technique for analyzing data corpus of narratives, 3) only mention data storytelling in the Abstract (e.g., as an application area of their research results) but do not study data stories in the main text, and 4) use data storytelling as a decorative term of data visualization or statistical report.
We ended up with \npublication~publications, including \npaper~papers, 10 books, and 5 book chapters. 


\subsection{Corpus Overview}
\label{sec:corpus_overview}
Here, we report the distributions of (i) contribution types, (ii) subject areas of venues, and (iii) application domains of the \npublication~publications in our corpus.

\textbf{Contribution types.} We classified the contribution types of the publications by adapting the six paper types of IEEE Visualization Conference (VIS) submissions. Their distributions are shown in Figure~\ref{fig:paper_overview} A.
Overall, most papers contributed \textit{Theoretical \& Empirical} knowledge to the foundation of data storytelling, with an increasing number of papers. We further observed a trend towards practical applications (\textit{Application} papers) and the development of systems and tools (\textit{Systems \& Tools} papers).

\textbf{Subject areas}. For subject areas, we classified venues based on the All Science Journal Classification Codes (ASJC) from Scopus~\cite{asjc2023}.
When Scopus provided multiple tags of the subject areas (e.g., art and humanities, and computer science) with ranks, we selected areas with the highest rank. 
We manually classified venues that were not included in Scopus into a category in ASJC by referencing the categories of similar venues and the venues' official websites to understand their major research topics and areas.
We observed a promisingly increased diversity of subject areas over time, with \textit{Engineering/Computer Science} leading the pack, as shown in Figure~\ref{fig:paper_overview} B.
There has been a rise in interdisciplinary research, with publications from various categories expanding. Specific fields, such as \textit{Social Science}, \textit{Arts and Humanities}, \textit{Medicine} (where most publications are about health communication), and \textit{Environmental Science} begin to claim a larger share of research.

\textbf{Application domains.} In our corpus, \napplication~publications explicitly mentioned that their research focused on specific application domains. 
Figure~\ref{fig:paper_overview} C presents the distribution of the application domains.
We identified 11 domains that had two or more publications and classified other domains into the \textit{Other} category.
\textit{Other} application domains include a diverse range, from cybersecurity and software to aviation, online video services, and beyond. 
The overview of the paper corpus underscores the expanding significance of data storytelling in shaping diverse facets of our increasingly data-dependent world, gaining cross-disciplinary academic interest in various application domains. 

\subsection{Data Analysis Approach}
In this section, we introduce the general collaborative coding process we took. The coders are the first, second, and last authors.
We applied an overlapping coding protocol that was used by previous work~\cite{burns2023we,mack2021we}. The coding process can be classified into three stages. 
In the first stage, the coders decided on a coding approach (e.g., thematic analysis and affinity map) and a coding framework. The coders then coded the same subset of publications and cross-checked everyone's coding results to clarify and refine the coding framework until reaching an agreement. 
In the second stage, following the coding framework, the coders divided the remaining data and coded it independently. At the middle point of the second stage, 15 publications were selected from every coder's coding results and cross-checked again to resolve any conflicts. The coding framework might have a few adjustments at this stage.
In the last stage, the coders finished the remaining coding tasks and met bi-weekly to discuss any problems encountered during coding. 

Next, we report two methods we applied to derive data story forms and analyze conceptualizations of data storytelling.
 
\subsubsection{Theme Analysis}
We extracted paragraphs and sentences from the papers and books in our corpus that explicitly described or defined what data storytelling is. At the beginning stage, the coding team applied an open coding scheme to the same set of 50 conceptualizations. In the first group discussion, the first author extracted and grouped concepts and descriptions from these conceptualizations. After several rounds of discussions, we refined the groups and generated codes and themes together. We then coded the rest of the conceptualizations. Overall, 16 codes and four themes were derived. Details are reported in Section \ref{sec:conceptualization}.

\subsubsection{Affinity Map Analysis}
Affinity map analysis was applied to both classify the forms of data stories and the conceptualizations.
For conceptualizations, we randomly selected 50 conceptualizations and had a group discussion where we iteratively grouped data items that took similar conceptualization approaches on the Figma whiteboard. Conceptualizations from the same group highlighted the same types of subjects, such as the goals and techniques of data storytelling. The classification ended when everyone looked through the conceptualizations in each group without proposing any changes. After that, we discussed the proper names and conceptualizations of each group. We concluded with six conceptualization types and labeled the types of the rest of the conceptualizations.
Data story forms were coded for every publication in the corpus. In the beginning, all coders independently coded the same set of 15 publications and applied the taxonomy of narrative visualizations from Segel and Heer~\cite{a1segel2010narrative}. However, new forms were found beyond the taxonomy. Thus, we included forms from other taxonomies~\cite{chen2023does, lan2023affective} as potential codes, and applied an open coding approach: the coders divided the remaining publications and labeled the forms either by selecting proper codes from past taxonomies or labeling freely. For example, we labeled forms like ``sketch infographics,'' ``AR,'' and ``light show.'' 27 different codes were generated. The codes were later grouped and merged into 12 types of forms using the affinity map approach.
 
\section{How Researchers Have Conceptualized Data Storytelling}
\label{sec:conceptualization}
This section reports our findings on how existing literature conceptualizes data storytelling.

\subsection{Themes Emerging in Conceptualizations}
\label{sec:theme}
We concluded four themes from the conceptualization analysis. 
They relate to the questions: What is being told in data storytelling? (\textbf{What}), What are the techniques used in data storytelling? (\textbf{How}), What effects should data storytelling achieve? (\textbf{Why}), and Who is the audience of data storytelling? (\textbf{Who}). Table \ref{tab:coding_summary} presents the counts of codes.

\begin{table*}[ht]
\centering
\renewcommand{\arraystretch}{1.3}
\begin{tabular}{|p{3cm}|p{11cm}|}
\hline
\textbf{Themes} & \textbf{Counts of Codes} \\
\hline

\textbf{What} &
None (47), Data Facts and Insights (101), and Beyond Data (39). \\
\hline

\textbf{How} &
None (26), Visuals (127), Narrative (104), Interaction (15), Others (10), and Data Science (9). \\
\hline

\textbf{Why} &
None (56), Inform and Explain (61), General Communication (47), Engage (27), Persuade (10), Evoke Emotions (8), Call-to-Action (7), Facilitate Retention (6), Provoke Thinking (6), Support Decision-making (5), Drive Change (4), Others (4) (Entertain (2), Facilitate Interaction (1), and Fulfill Business Needs (1)). \\
\hline

\textbf{Who} &
None (122), General (35), and Specific (15). \\
\hline

\end{tabular}
\caption{The counts of codes in the \nconceptualization~conceptualizations of data storytelling under the What, How, and Why, and Who Themes.}
\label{tab:coding_summary}
\end{table*}


\subsubsection{What Is Being Told in Data Storytelling?}
\label{sec:content}
This theme was applied to \textbf{123/\nconceptualization}~conceptualizations, derived from the descriptions in these conceptualizations that stated what data storytelling should communicate. There were two codes, \textit{Data} and \textit{Beyond Data}, under this theme. The first code \textbf{Data Facts and Insights (101)} was applied to conceptualizations that stated data storytelling was grounded in data. 
They argued that data storytelling should convey data, insights from data analysis, or even the data analysis process (e.g., \cite{a138wilkerson2021reflective}). The second code \textbf{Beyond Data (39)} was applied to conceptualizations that argued data storytelling concerned higher-level messages beyond data, such as \emph{thought-provoking opinions}~\cite{lee2015more}, \emph{specific questions}~\cite{a189pozdniakov2023teachers}, and \emph{intended stories}~\cite{hullman2011visualization}. 84 conceptualizations only discussed data-related information. 22 emphasized higher-level messages. 17 highlighted both data and beyond-data messages.

\subsubsection{What Are the Techniques Used in Data Storytelling?} 
\label{sec:technique}
\textbf{144/\nconceptualization}~conceptualizations described the techniques for data storytelling. We categorized them into the following.

\textbf{Visual (126).} In conceptualizations that mentioned visual techniques, we observed different scopes of visual techniques. 45 conceptualizations mentioned \emph{chart} (e.g.,~\cite{a189pozdniakov2023teachers,a120chokki2023conventional,cao2020user}), \emph{data visualization} (e.g.,~\cite{a116metoyer2018coupling,a68bradbury2020documentary}) or \emph{information visualization} (e.g.,~\cite{a369obie2020authoring,a123wang2021auto,a108echeverria2018driving}).
78 conceptualizations used general terms like \emph{visualization} (e.g.,~\cite{a252whitlock2020hydrogenar,a25cao2023dataparticles,a18lan2021kineticharts}), and \emph{visuals} (e.g.,~\cite{a101ojo2018patterns,a130janowski2019mediating,a38martinez2020data}), indicating that data storytelling could include visual techniques other than data visualizations.  
Some conceptualizations mentioned concrete visual forms except data visualizations. For instance, Bentley et al.~\cite{bentley2022feasibility} defined: \emph{Narrative visualization (NV) is a technique that uses drawings, photographs, and text to contextualize data.}   
The most discussed visuals were \emph{Text} (17) and \emph{Annotation} (10). This reflects that the combination of text and data visualizations was the most commonly discussed form of data storytelling. 
Other visual forms that were mentioned more than once included \textit{Animation} (4), \textit{Illustrations} (3),  \textit{Photographs} (2), \textit{Image} (2), \textit{Drawings} (2), \textit{Storyboards} (2).

\textbf{Narrative (104).} Most of the conceptualizations (77 out of 104) that include narrative techniques used general terms such as \emph{narrative techniques} (e.g.,~\cite{a39concannon2020brooke,m7zhao2023stories,a10choudhry2020once}) and \emph{storytelling techniques} (e.g.,~\cite{a369obie2020authoring, a18lan2021kineticharts, a481renda2023melody}).
Some conceptualizations included more specific techniques.
The most frequently mentioned one was the structure of data stories (16). 
It did not refer to traditional narrative structures such as \emph{Freytag's Pyramid}~\cite{freytag1895technique} or \emph{Hero's Journey}~\cite {campbell2008hero}. These conceptualizations describe that data storytelling should have a storyline, organized story pieces, or connected data visualizations.
Another technique mentioned more than twice was \emph{Rhetorical Techniques (6)}.
Five conceptualizations referred to rhetorical techniques as what data storytellers would consciously or sub-consciously apply to frame the audience's reception of the messages for the purpose of persuasion, which was first introduced to narrative visualization by Hullman and Diakopoulos~\cite{hullman2011visualization}. One conceptualization (\emph{The authors define narrative visualization as...that engage the viewer with metaphor and storytelling}~\cite{a209fry2013teaching}) highlighted a specific rhetorical device --- metaphor.

\textbf{Interaction (15).} This code derived from conceptualizations that used terms like \emph{interactive graphics}(e.g., \cite{zheng2022evaluating}) and \emph{interactive data visualizations} (e.g., \cite{a302bohman2015data}) or descriptions like \emph{...through data storytelling, in the form of charts and interactive interfaces...}~\cite{shan2022research}.  

\textbf{Data Science (6).} Six conceptualizations discussed the premise of data storytelling about the application of data science. Particularly, they stated that data storytelling combined or employed \emph{data analysis and modeling methods}~(e.g., \cite{a489daradkeh2023data}), \emph{data science}~(e.g., \cite{a123wang2021auto}), and \emph{data analysis}~(e.g., \cite{a493chen2023embedding}). 

\textbf{Others (10).} This code represents techniques that were mentioned only once or were ambiguous. For example, three conceptualizations described \emph{narrative visualization} as the technique of data stories~\cite{a24sun2022erato,a12shi2020calliope,a474wu2023socrates} (e.g., \emph{A visual data story is a series of connected data facts shown in form of a narrative visualization...}~\cite{a24sun2022erato}), suggesting that narrative visualization is a subset of data story. Some conceptualizations mentioned \emph{embedding mechanisms}~\cite{a116metoyer2018coupling, anupama2020narrative} and \emph{compression techniques}~\cite{aydin2020data}.

\subsubsection{What Effects Should Data Storytelling Achieve?} 
\label{sec:objective}
\textbf{114/\nconceptualization}~conceptualizations described the effects data storytelling should achieve.
Some conceptualizations suggested that data storytelling served for \textbf{General Communication (47)} goal. They provided general descriptions such as to tell or convey a \emph{story(-ies)} (e.g.,~\cite{a267wirfs2021examining, a17yang2021design}), \emph{present data} (e.g.,~\cite{a267wirfs2021examining, a17yang2021design}), and \emph{present particular perspectives}~\cite{a116metoyer2018coupling}.  
More conceptualizations specified the communication goals.
\textbf{Inform and Explain (61)} was the dominant goal in these conceptualizations. These conceptualizations highlighted a basic requirement of data storytelling: making the findings from data understandable to the audience. This aligned with the primary objective of data visualization, which is to provide a human-friendly interface for understanding data, which implies an objective presentation of data. 
Some conceptualizations indicated that data storytelling aims further, seeking not only to inform but also to \textbf{Persuade} \textbf{(10)}. More specific goals were \textbf{\textbf{Provoke Thinking (6)}, Call-to-Action} \textbf{(7)}, \textbf{Drive Changes} \textbf{(4)}, and \textbf{Support Decision-making (5)}.
Furthermore, data storytelling was expected to \textbf{Engage (28)},  \textbf{Evoke Emotions (8)} and \textbf{Facilitate Retention (6)}. \textbf{Others (4)} goals that were mentioned less than three times included \emph{fulfill  specific  business  needs}~\cite{a489daradkeh2023data}, \emph{reach a wide or targeted audience}~\cite{bc1isenberg2018immersive} and \emph{entertaining the viewers}~\cite{lee2015more}.

\subsubsection{Who Is the Audience of Data Storytelling?} 
\label{sec:audience}
Most conceptualizations did not concern who the audience of data storytelling was, but \textbf{48/\nconceptualization}~conceptualizations.
Nonetheless, we found nuanced differences in their description of the audience and identified two categories. 
Conceptualizations in the category \textbf{General (35}) used general terms like \emph{general audience} (e.g., ~\cite{a401meuschke2022narrative, a448zhang2022visual}), \emph{user}~\cite{cao2020user}, \emph{readers}~\cite{a40heyer2020pushing} and \emph{recipients}~\cite{a278lindoo2022case}. Other conceptualizations in the category \textbf{Specific (15)}, on the other hand, suggested that data storytelling should have the \emph{target audience} (e.g., ~\cite{a130janowski2019mediating,a101ojo2018patterns,a489daradkeh2023data}) and the audience could be \emph{non-technical audience} (e.g.,~\cite{a119brolchain2017extending,a120chokki2023conventional,a481renda2023melody}) or \emph{decision makers}~\cite{a112behera2019big}.








\subsection{Five Types of Conceptualizations}
We identified five perspectives researchers took to conceptualize data storytelling, namely the \textit{Technique-and-goal perspective (D1)}, \textit{Visualization-centric perspective (D2)}, \textit{Technique-centric perspective (D3)}, \textit{Goal-centric perspective (D4)}, and \textit{Interdisciplinary perspective (D5)}. 
These perspectives have various focuses on the themes and codes in Section~\ref{sec:theme}. 

\textbf{D1. Technique-and-goal perspective (78/\nconceptualization).} This type of conceptualization took around 50\% of all conceptualizations. The authors defined data storytelling by pointing out its core objectives
and techniques. For example, Stanislav et al.~\cite{a189pozdniakov2023teachers} defined that \emph{Data storytelling (DS) is an information compression technique that can be applied to help the audience focus on responding to specific questions by combining charts, text, and other resources to emphasise the data points and the evidence that are more relevant to such questions.} Lan et al.~\cite{a45lan2022negative} described that \emph{Data stories, as a visual form that combines data visualization with narratives, interaction, embellishment, or animation, are increasingly embraced by content producers and disseminators to communicate data to viewers.}  

\textbf{D2. Visualization-centric perspective (23/\nconceptualization).} 
Unlike other conceptualization types where no single term (e.g., \emph{data story}, \emph{data storytelling}, \emph{narrative visualization}) was used dominantly, the majority of these conceptualizations (15) employed the term \emph{narrative visualization}.
The authors defined data storytelling (mostly \emph{narrative visualization}) as a particular data visualization type that integrates storytelling techniques into its design or as a visual form mainly consisting of data visualizations. 
They explicitly suggested that data visualization was its essence.
Examples are \emph{Narrative visualizations are data visualizations with embedded `stories' presenting particular perspectives using various embedding mechanisms.}~\cite{a116metoyer2018coupling}) and \emph{Narrative visualization is a form of data storytelling where statistical graphics are used to complement (or even replace parts of) a story}~\cite{bryan2020analyzing}.
As the term \emph{narrative visualization} was interchangeably used by many papers with other terms like \emph{data story} and \emph{data storytelling}, it is hard to tell if these conceptualizations consider data storytelling with a relatively narrow scope compared with other conceptualizations, or if they regard \emph{narrative visualization} as a term independent from \emph{data storytelling}.
This finding indicated that more nuanced terms are desired to cover different scopes of data storytelling.

\textbf{D3. Technique-centric perspective (26/\nconceptualization).} The authors defined data storytelling by mainly describing components and techniques that build data stories. Some of them mentioned a vague goal of data storytelling, such as \emph{telling stories}.
Examples include \emph{We define a narrative visualization as a set of visualization components, control widgets and annotations that are coordinated by a narrative state machine}~\cite{a65satyanarayan2014authoring} and \emph{...a data story can be defined as the outcome of employing storytelling techniques that integrate data visualization elements within the narrative. This combination results in a cohesive presentation of textual content and visual elements, such as charts, arranged in a logical sequence determined by the author}~\cite{a481renda2023melody}.

\textbf{D4. Goal-centric perspective (15/\nconceptualization).} The authors defined data storytelling by focusing on the goals data storytelling should achieve. For example, Brolcháin et al.~\cite{a119brolchain2017extending} defined that \emph{DDS \textit{(note from the authors: Data-driven storytelling)} can be explained as a process of translating data analysis into simple, logical stories that can be understood by a non-technical audience}. Berendsen et al.~\cite{a212berendsen2018digital} described that \emph{The terms 'data storytelling' or 'data-driven storytelling' have been used to refer to how insights are uncovered from data and communicated to others effectively so that they can be translated into actions or changes}.

\textbf{D5. Interdisciplinary perspective (18/\nconceptualization).} The authors defined data storytelling in a high-level manner by emphasizing it as the combination of visualization and storytelling techniques.
For example, Choudhry et al.~\cite{a10choudhry2020once} stated that \emph{Visualization, inherently, is inclined for communication by virtue of its graphical form, resulting in the notion of communication-minded visualization. Combining the idea of communication-minded visualization with storytelling yields the notion of data-driven storytelling: narrative techniques for data}.
Obie et al.~\cite{a369obie2020authoring} described \emph{Narrative visualisation (visual data stories), i.e., the combination of information visualisation with storytelling mechanisms...}
These conceptualizations did not explain how the techniques from the two fields were combined or indicate which one plays a dominant role.

\begin{table*}[t]
\label{tab:form}
\centering
\renewcommand{\arraystretch}{1.35}
\setlength{\tabcolsep}{10pt}

\begin{tabular}{|
  >{\centering\arraybackslash}p{0.23\textwidth}|
  >{\centering\arraybackslash}p{0.2\textwidth}|
  >{\centering\arraybackslash}p{0.2\textwidth}|
  >{\centering\arraybackslash}p{0.2\textwidth}|
}
\specialrule{1.2pt}{0pt}{0pt} 
\textbf{Our work} & \textbf{Segel \& Heer \cite{a1segel2010narrative}} & \textbf{Chen et al. \cite{chen2023does}} & \textbf{Lan et al. \cite{lan2023affective}} \\
\specialrule{1.0pt}{0pt}{0pt} 

Data Comics
& Comic Strip 
& Data Comics 
& \multirow{5}{*}{Static image/painting}
\\ \cline{1-3}

\multirow{3}{*}{Infographic}
  & Annotated Chart
  & Annotated Chart
  & 
\\ \cline{2-3}
  & Flow Chart
  & Timeline \& storyline
  &
\\ \cline{2-3}
  & Partitioned Poster
  & \multirow{2}{*}{Infographics}
  &
\\ \cline{1-2}

\multirow{2}{*}{Multimedia Article}
  & Magazine Style
  &
  &
\\ \cline{2-4}
  & Slideshow
  & Scrolllytelling \& Slideshow
  & \multirow{3}{*}{Interactive Interface}
\\ \cline{1-3}

Dashboard \& System 
& - 
& - 
& 
\\ \cline{1-3}
XR Storytelling & - & - & \\ \hline

Oral Presentation & - & - & - \\ \hline
Video & Film/video/animation & Data Videos & Video \\ \hline
Artifact & - & - & Artifact \\ \hline
Installation & - & - & Installation \\ \hline
Event & - & - & Events \\ \hline
Sound & - & - & - \\ \hline
Text & - & - & - \\ 

\specialrule{1.2pt}{0pt}{0pt} 
\end{tabular}
\caption{The table presents the relationship of our taxonomy with past taxonomies in data storytelling that share the same or similar forms.}
\end{table*}

\section{Data Storytelling Forms}
\label{sec:forms}
Extending on taxonomies of past data storytelling surveys and taxonomies \cite{a1segel2010narrative,chen2023does,ZhaoElmqvist2023StoriesWeTell}, we identified the following common forms of data stories that have been applied in existing research. Table
\ref{tab:form} presents the relationships of forms in our taxonomy with those from past surveys.
The taxonomy of forms presents how data stories are implemented with different media techniques and designs. It provides a quick query and concrete images of what a data story can appear like.
In addition to forms, we observed some basic attributes that distinguish the storytelling experiences from both narrative perspectives and human-computer interaction perspectives. Those attributes can be combined with different data story forms. 

\begin{figure*}
  \centering
  \includegraphics[width=0.87\textwidth]{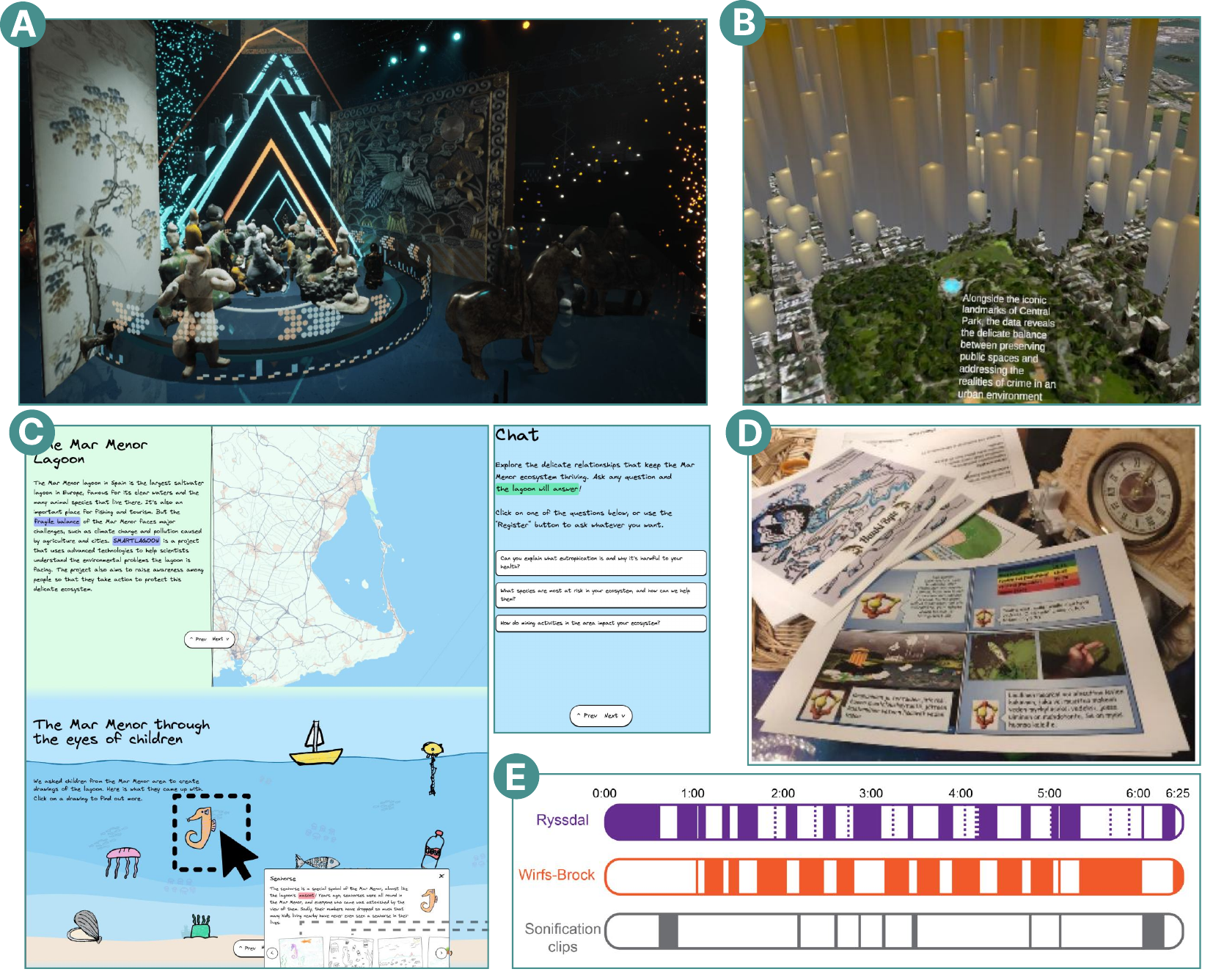}
\caption{Examples of data stories with particular forms: (A) Installation: A stage show for stories about cultural heritage, where the lyrics and visual effects encode data \cite{a550xu2025artifact}. (B) XR Storytelling: A VR story in a 3D environment from \cite{zhang2025st}. (C) Dashboard \& System: An interactive system that integrates narrative visualizations and an AI-driven conversational interface \cite{a542tumedei2025drawing}. (D) Event: A participatory workshop with a board game for educational purposes. (E) Audio: A radio data story that integrates data sonification clips \cite{a518kauer2025towards}.}
  \label{fig:case_form}    
\end{figure*}

\subsection{Distributions of Forms}
Most of the forms we identified are well-established or emerging forms that have already attracted considerable research attention. Not all cases in our corpus involve a concrete data story form. Some of them are about general techniques or theories that can be applied to any form. 
The most common forms are \textbf{Multimedia Article (103)}, \textbf{Infographic  (53)}, and \textbf{Video (53)}. The multimedia article form is mostly common in online news articles. It includes the narrative visualization genres \textit{Magazine Style} and \textit{Slideshow} from the taxonomy by Segel and Heer \cite{a1segel2010narrative}, as well as \textit{Scrollytelling}, a scroll-driven long-form format in which textual and visual elements appear or change as the audience scrolls through an online article \cite{seyser2018scrollytelling}.
While the \textbf{Data Comics (17)}, \textbf{\textbf{XR Storytelling (17)}}, and \textbf{Oral Presentation (9)} forms are less common in our corpus, they have already been discussed and drive standalone research streams. For instance, research has identified design patterns and conducted evaluations of data comics \cite{bach2019design,wang2022interactive, wang2021reporting} and immersive data-driven storytelling with XR techniques \cite{bc1isenberg2018immersive,mendez2025immersive,yang2023understanding}. Oral presentations are discussed as an important scenario of data storytelling \cite{Kosara2013,ZhaoElmqvist2023StoriesWeTell} with tools developed for authoring (a-)synchronous presentations \cite{takahira2025tangiblenet,hall2022augmented}. 

On the contrary, the following data story forms are under-explored with little or no research that has derived or studied their design patterns. An \textbf{Artifact (3)} uses physical objects to encode data and present a story. The idea is closely related to data physicalization~\cite{jansen2015opportunities}.
An \textbf{Installation (7)} tells data stories through large-scale, mixed-media constructions in a specific physical space (see an example in \autoref{fig:case_form} A).
In a data story in the form of \textbf{Event (5)}, the audience participates in activities, such as in a board game \cite{a53hasan2022playing} (\autoref{fig:case_form} D). They are invited to act as part of the story, and sometimes their reactions and choices decide how the story progresses. While they are less discussed, a survey on affective data visualizations found that those forms have already been applied to evoke emotions~\cite{lan2023affective}. 

Finally, we identified several unexpected forms. In particular, \textbf{Text (7)} and \textbf{Sound (2)} are notable because data stories are typically associated with visual presentation, as revealed by the analysis of data storytelling conceptualizations. An example is given in \autoref{fig:case_form} E, where data sonification clips are integrated into a radio data story \cite{a518kauer2025towards}. A widespread assumption indicated in those conceptualizations is that visual representations are central to the appeal and effectiveness of data storytelling. This underscores that data stories can also be constructed without visuals. Rather than diminishing the power of data stories, the use of text and sound could broaden their accessibility, for example, by better supporting audiences with visual impairments or enabling consumption in eyes-free contexts \cite{a518kauer2025towards}.
Last but not least, some data stories are presented through interactive \textbf{Dashboard \& System (33)}, which presents a nonlinear space of multiple possible views and interactions, and the narrative emerges from the audience’s exploration rather than a linear sequence, as an article normally does. A system normally involves complex data---high-dimensional, large-scale, or heterogeneous. An example is given in \autoref{fig:case_form} D. Sometimes, data stories are embedded as part of a system to onboard the audience or highlight key messages. 

\subsection{Dimensions of the Audience Experiences in Forms}

Typologies of forms can help name recurring ways in which data stories are presented, yet using the typologies solely to describe data stories introduces challenges. First, the literature and practice often mix form, format, and genre, which can blur what exactly is being categorized. Second, category boundaries are porous: the same artifact can legitimately fit multiple types (e.g., an infographic can adopt a chunked, panel-like composition that reads similarly to a comic). Third, typologies struggle to account for hybrid and newly emerging presentations.

Given these challenges, we additionally characterize data-story presentations using four spectrum-based dimensions that capture experience-shaping properties at a more abstract level than forms. These dimensions include: 1) spatiality, ranging from 1D to 3D; 2) sensory modalities, ranging from unimodal to multisensory; 3) ordering, ranging from ordered progression to free-form exploration; and 4) audience co-authorship, ranging from low to high. We illustrate each dimension with examples and discuss how it shapes data storytelling experiences.

\subsubsection{Spatiality}
\noindent\includegraphics[width=\columnwidth]{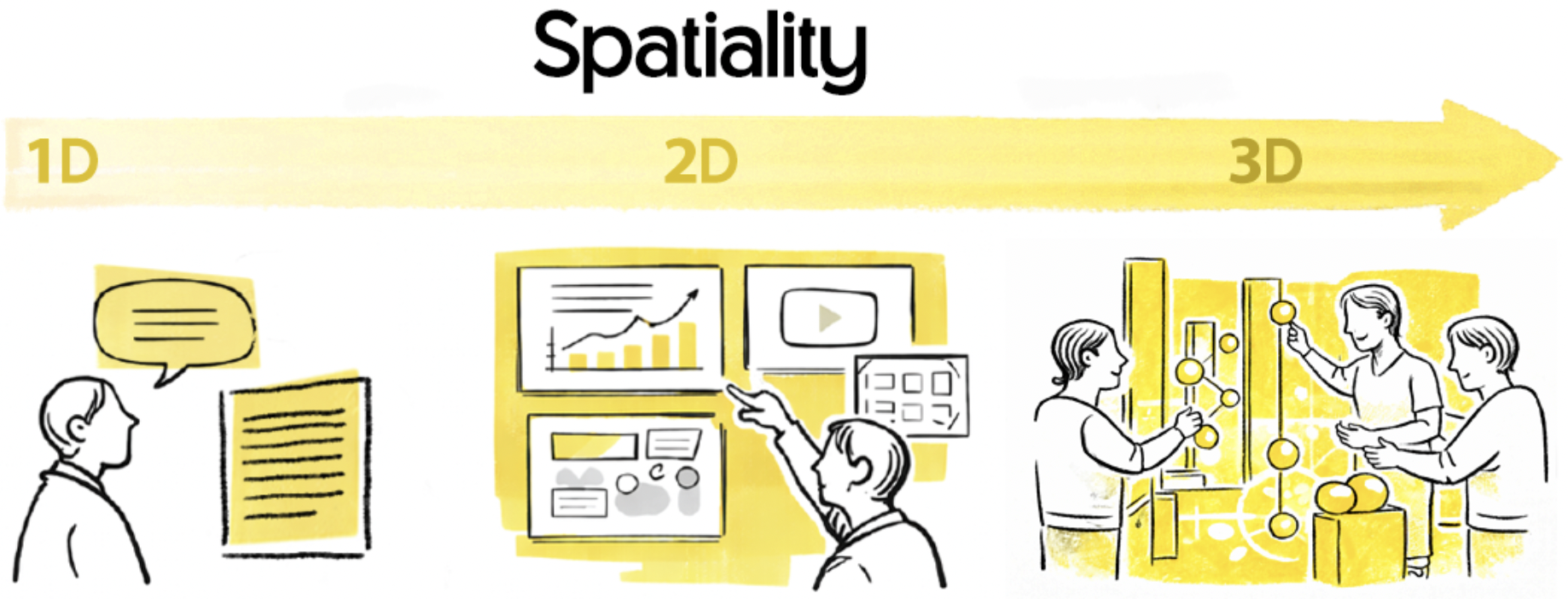}
This dimension concerns the spatial experience in which the audience approaches a data story. A small portion of data stories in our corpus are experienced in a 1D space, primarily through pure text or audio. 2D data stories encompass forms such as data comics, infographics, multimedia articles, videos, and systems. Additionally, we have observed some data stories in 3D environments, such as immersive data stories based on AR/VR technologies (\autoref{fig:case_form} B), physical installations (\autoref{fig:case_form} A), live oral presentations and events, as well as fabricated artifacts. 
As the spatial dimension increases, the modalities the audience can experience expand: 2D introduces imagery information compared to 1D, while 3D typically incorporates bodily interaction and physical materiality.

Is more spatial experience necessarily better? One argument contends that since humans inherently interact with the world through multimodal and bodily means, richer spatiality in data stories leads to more environmental and context-embedded presentation and interaction~\cite{jiang2020baguamarsh,ren2018xrcreator,liu2021narrative,luo2024between}, thereby enhancing immersion and the sense of presence~\cite{jiang2020baguamarsh,luo2024between,jain2025strollytelling,zhu2024reader}. A spatial experience is also said to be able to leverage the audience’s kinesthetic intelligence~\cite{bhargava2022data} and facilitate emotional engagement and reflection through embodied interaction~\cite{desjardins2019listeningcups}. 
However, for the purpose of data understanding, several studies argued that 3D data stories can aid in understanding complex data that might otherwise be compressed or lost in printed or desktop media~\cite{dibenigno2021flow,liu2021narrative,zhang2025st}.
In narrative terms, Hancox~\cite{hancox2021revolution} offered a further perspective, suggesting that mixed-reality stories can create a space where \emph{we as humans are not necessarily the centre of these stories; rather, we are encouraged to view ourselves as part of a wider ecosystem.} This shift, he argued, can foster deeper exploration of non-human perspectives and engagement with collective issues such as global warming. 
Additionally, some scholars have noted that richer spatiality can provide greater narrative context for data~\cite{a486whitlock2020hydrogenar}, for example, by displaying data stories in public spaces to facilitate engagement with civic data~\cite{claes2017impact}.

However, another line of thinking suggests that 1D and 2D forms also have their unique merits. For example, while 3D representations excel at conveying spatial-related data such as height, volume, and speed~\cite{zhang2025st,zhu2024reader}, they can introduce perceptual challenges and may not always outperform their 1D and 2D counterparts~\cite{zhou2023data}. Also, the development of AR/VR applications, physicalizations, and installations is often resource-intensive, requiring significant time and material investment.
In fact, from a broader media-evolution standpoint, 2D forms such as images and short videos are still the dominant media today; besides, the unique appeal of 1D formats like podcasts endures. For example, researchers~\cite{mchugh2022power} have found that podcasts strategically employ an \emph{absence} or \emph{negative space} in spatiality. This compels the brain to actively construct meaning, scenes, and emotional resonance internally while also freeing cognitive resources to allow for parallel processing (e.g., listening while commuting).
Echoing this, a few works in our corpus have employed data sonification for storytelling to foster a distinct form of deep, reflective engagement~\cite{a267wirfs2021examining}, which we think constitutes an interesting and complementary direction to the significant pursuit of constructing rich spatial experiences.


\subsubsection{Sensory Modalities}
\noindent\includegraphics[width=\columnwidth]{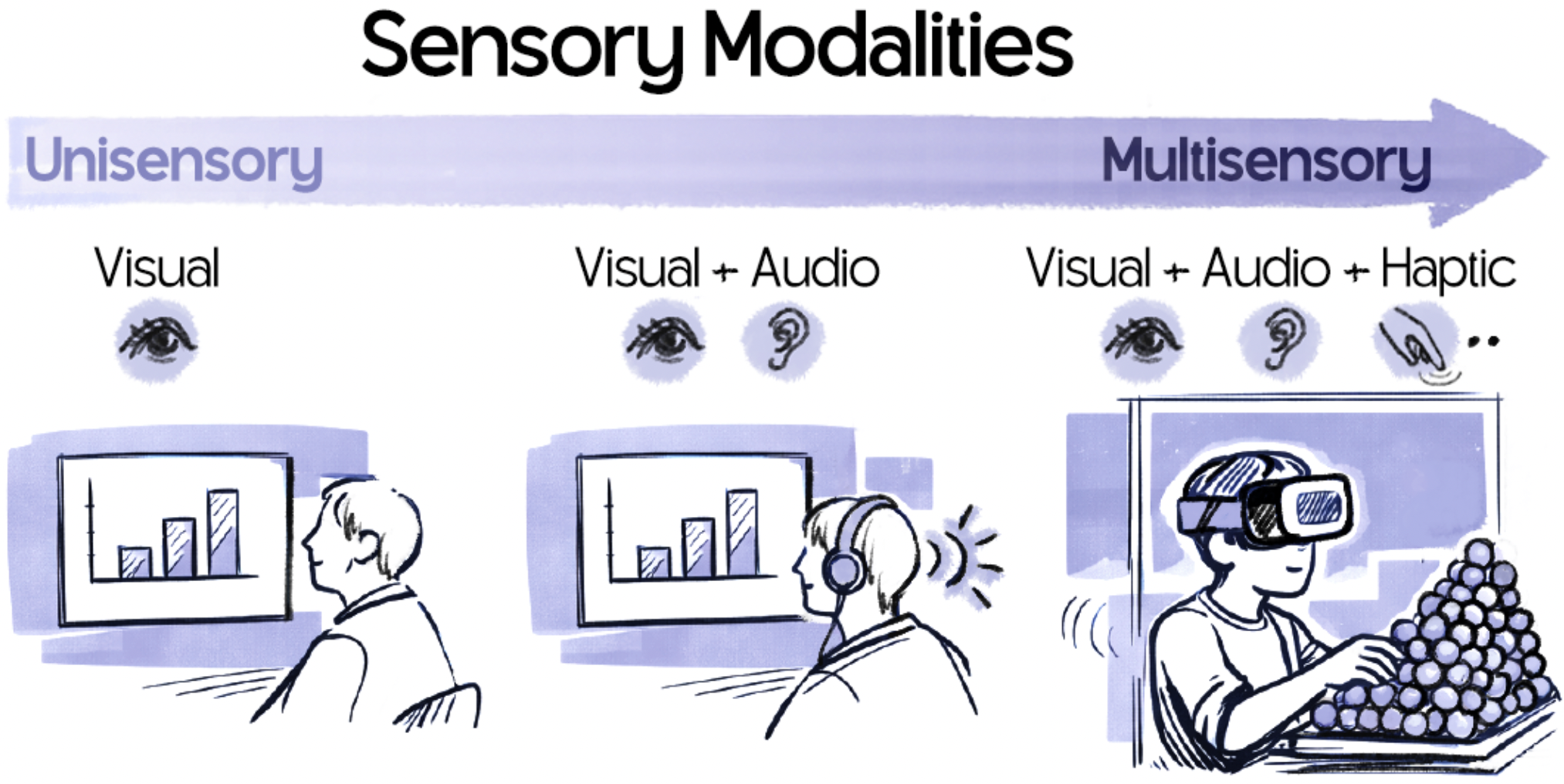}
This spectrum involves the range of sensory modalities a data story leverages to present information, from unimodal text to multisensory experiences (visual, auditory, haptic, olfactory).
The sensory-modality spectrum captures how data stories move from unimodal visual or text to multisensory experiences (e.g., visual, auditory, haptic). It is not surprising that most prior data storytelling research focuses on visual (often with supporting text) and on techniques for visual communication, since dominant forms like articles, infographics, and videos are built around screens. However, the less common forms in our taxonomy, especially \textbf{Sound} and \textbf{Artifacts}, remind us that data stories can also be told through hearing and touch, not only sight.
Looking forward, combining modalities may offer benefits beyond redundancy: sound and music can convey tone and emotion in ways that are difficult to achieve with charts alone \cite{xu2022wow}. Haptics and physical materials can support embodied engagement, which can be incorporated into stories within a 3D environment. At the same time, multisensory storytelling introduces design challenges, such as coordinating timing and meaning across channels and avoiding sensory overload. For example, research in data videos has explored design guidelines and tools for coordinating animations and narrations \cite{shen2023data,shen2025reflect}, while other possible combinations of different sensory modalities remain less-explored.

\subsubsection{Ordering}
\noindent\includegraphics[width=\columnwidth]{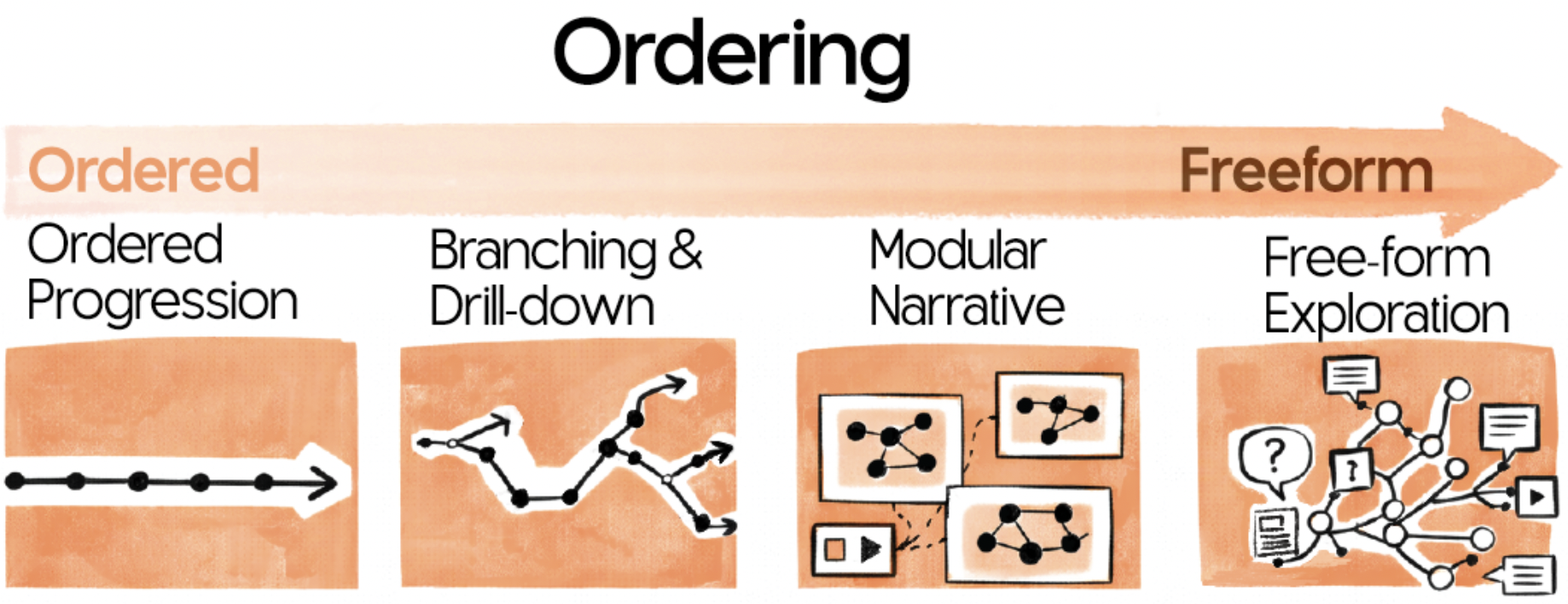}
Ordering in data stories can be described as the degree to which the ordering of experience is predetermined by the author, versus assembled on the fly by the audience through exploration. At the ordered end of this spectrum, progression is structured through fixed sequencing mechanisms such as fixed time (a chronological backbone that implies a stable next state), fixed path (a prescribed spatial route that encodes narrative order~\cite{jain2025strollytelling,zhang2025st}), or fixed task flow (a stepwise set of screens, scenes, or actions that the audience is expected to follow~\cite{a252whitlock2020hydrogenar}). 

Moving from 1) ordered progression toward 4) free-form exploration, ordering becomes less prescribed and more conditional, selective, or audience-constructed. Exploration has been found using 2) branching and drill-down (different next steps depending on choices), for example in interactive data comics where structure operations such as append, replace, and load layout can introduce new panels, swap panels, or switch to an alternative layout, thereby changing the set and potential order of panels shown and enabling branching paths through the story~\cite{wang2022interactive}. 3) Modular narrative (jumping between narrative modules without a required sequence); we borrow the concept from film study that articulates a sense of time as divisible and subject to manipulation~\cite{cameron2008modular}. For example, the learner-facing data stories interface where different data stories are accessed via a set of buttons and learners navigate one data story at a time, allowing them to choose which story module to view next~\cite{fernandez2021storytelling}. 4) Freeform exploration (open-ended questions or exploration with minimal authorial gating), exemplified by the Mar Menor system’s conversational stage, where the audience can ask questions to the AI persona of the lagoon and thus construct their own trajectory through the content via inquiry rather than preauthored sequencing~\cite{Tumedei25Drawings}. Ordered progression can be useful when storytellers need steady pacing and shared takeaways. Open exploration helps when sensemaking depends on comparison and personalization.

\subsubsection{Audience Co-Authorship}
\label{sec:co-authorship}
\noindent\includegraphics[width=\columnwidth]{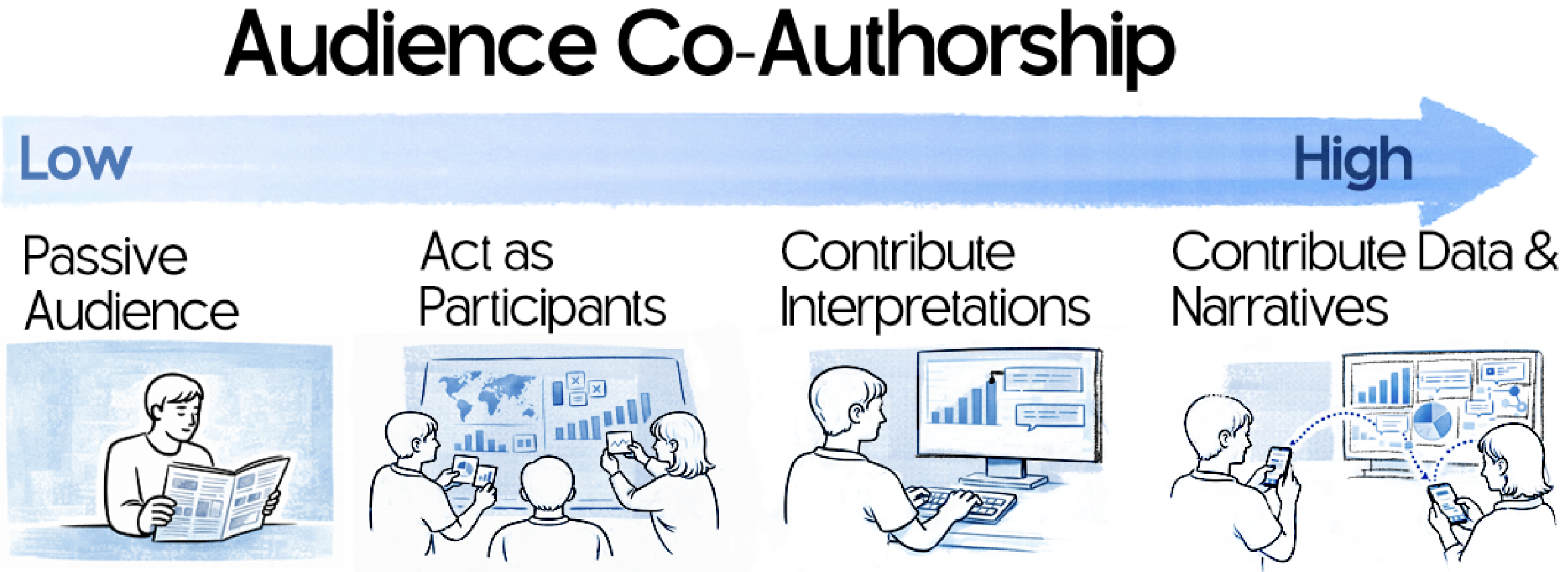}
While the ordering concerns between the audience and authors, who and how much they can control the ordering of the storytelling experiences in time and space, co-authorship decides who and how much they contribute to the story content.
There are several ways that the audience can co-author. For example, in our corpus, we observed that the audience can: 1) act as participants within the data story (e.g., playing as a character in a card game made with data comics \cite{a53hasan2022playing}), 2) contribute interpretations of the data (e.g., annotations in charts from the audience shared among each other \cite{a518kauer2025towards}), and 3) contribute personal data and narratives to the story (e.g., responding to a live survey the data of which become part of narrations in a live public debate on television \cite{a129robinson2014storied}).

When the audience co-authors, we expect several benefits. Most immediately, engagement tends to increase, as participants can feel a sense of agency in shaping the story. Co-authorship should enhance the relevance of the narrative to the audience, as their contributions reflect their experiences, perspectives, and interpretations. Furthermore, involving multiple participants can enrich the story with a broader range of viewpoints, potentially uncovering insights or interpretations that the original storytellers might not have anticipated. Creating stories that allow for audience co-authorship fundamentally shifts the role of the storyteller. Rather than driving the narrative in a top-down manner, storytellers must relinquish some control over the attention of the audience, their interpretation of the data, and the direction of the story. In other words, the narrative may evolve in ways that diverge from the storytellers’ original intentions or desired outcomes. This introduces both a creative opportunity and a challenge: stories become collaborative, emergent spaces where authority is negotiated, and meaning is co-constructed rather than prescribed.

\section{Comparing Data Storytelling with Classic Narratology}
\label{sec:concepts}


The term ``narrative,'' frequently cited as integral to data storytelling, is broadly utilized in narratology yet lacks a precise conceptualization in the literature we examined. 
This gap prompts an essential inquiry: How do the elements of a data story align with traditional narrative components? This understanding promises to expand the storytelling toolkit, facilitating the melding of timeless narrative charm with the nuances of data communication. 
Through a focused review of narratology literature, we delve into the adaptation of three fundamental narrative elements—\textit{Event}, \textit{Character}, and \textit{Plot}—in data storytelling. 
In this section, we juxtapose classic narratological theories with conceptualizations of data storytelling from our review and insights from research within our corpus, shedding light on the interplay between traditional narrative structures and data storytelling practices.

\begin{figure*}
  \centering
  \includegraphics[width=\textwidth]{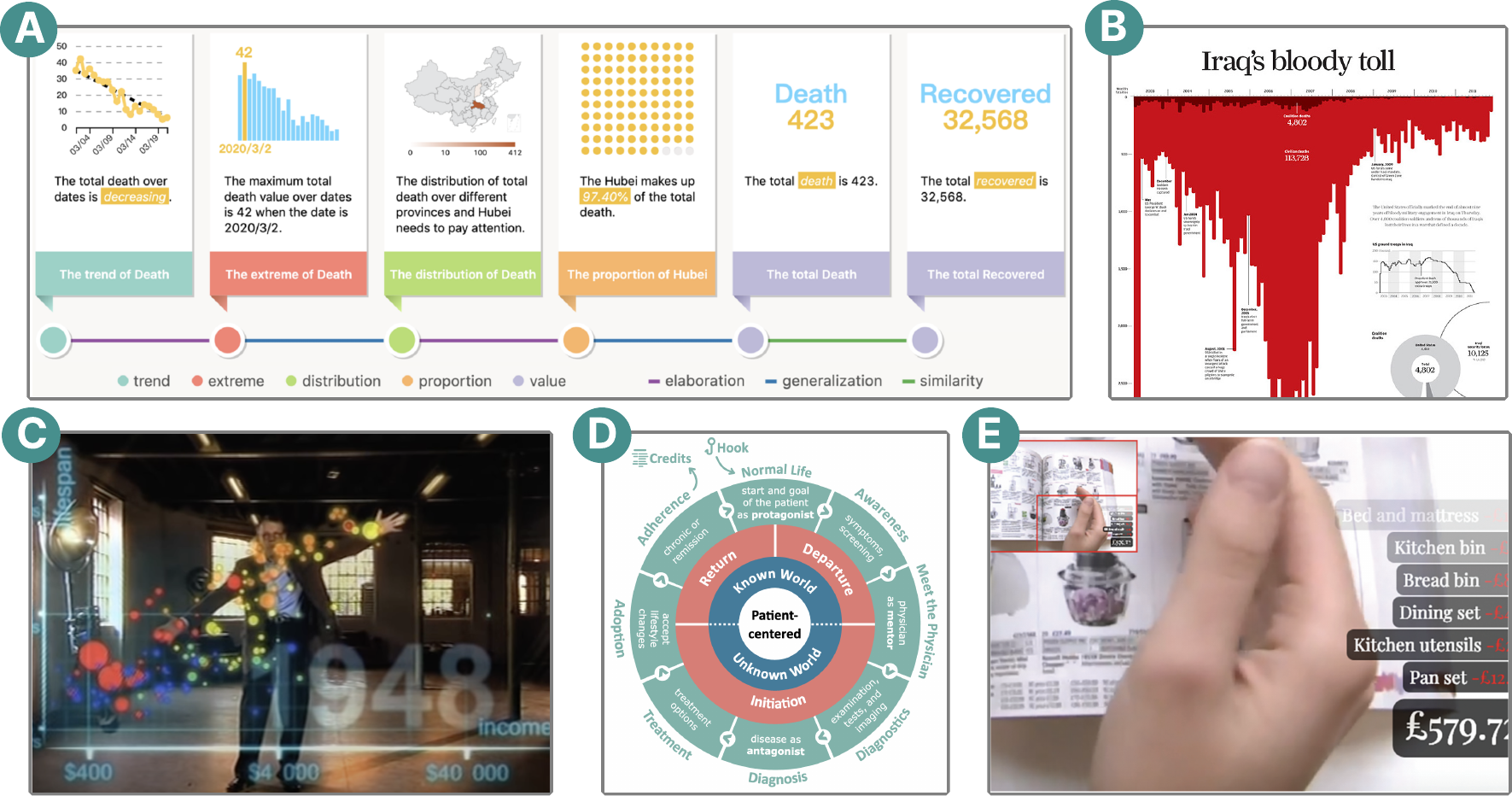}
  \caption{Examples of data stories with different adaptations of classic narrative elements: (A) a data story generated by Calliope~\cite{a12shi2020calliope}, where the narrative plot is based on the data relations of data facts. (B) the story \textit{Iraq's Bloody Toll}~\cite{iraq2011} with an abstract ``data character'' labeled by~\cite{a472dasu2023character}, (C) the story \textit{200 Countries, 200 Years, 4 Minutes} with a narrator shown in the camera, (D) a data story plots design that applies the Hero's Journey structure to tell disease data~\cite{a476mittenentzwei2023disease}, and (E) the story \textit{Brooke Leave Home} with an imaginary protagonist to tell open data about the care system in England~\cite{a39concannon2020brooke}.
}
  \label{fig:case}    
\end{figure*}

\subsection{Event}
\textit{Narratological perspectives.} 
In a thorough review of what a narrative is, H. Poter Abbott  \cite{abbott2020cambridge} proposed a minimum conceptualization: \emph{Narrative is the representation of an event or a series of events.} 
He further pointed out that most narrative theorists consider narrative should at least have more than one event, narrative events should have causal-and-effect relationships, or a narrative must involve human or humanlike characters. 
Undoubtedly, \emph{Event} is a critical feature that distinguishes narrative from other forms of writing~\cite{abbott2020cambridge}.
It is commonly agreed that a narrative event must change the state of characters or the broad storyworld~\cite{abbott2020cambridge}. 
For example, Mieke Bal defined \emph{events as the transition from one state to another state, caused or experienced by actors}~\cite{bal2009narratology}. Seymour Chatman stated that \emph{Events are either actions (acts) or happenings. Both are changes of state. An action is brought about by an agent or one that affects - a patient}~\cite{chatman1978story}.
Chatman gave a mini plot as an example: \emph{(1) Peter fell ill. (2) He died. (3) He had no friends or relatives. (4) Only one person came to his funeral.} He explained that statements (1), (2), and (4) depict events, while (3) is only a description~\cite{chatman1978story}.
The change of status in a narrative event is important for pushing forward the plot, indicating to what degree a protagonist achieves its goals, or surprising and evoking the emotions of the audience.

\textit{Data storytelling perspectives.} 
\textbf{Data storytelling research does not strictly distinguish \emph{description} from \emph{event}.} As discussed in Section~\ref{sec:content}, many conceptualizations indicate that data storytelling is only about the conveying of data. This indicates an equal role of data findings with \emph{narrative events}. A typical example is the data stories generalization tools in recent years. Those tools generate and organize different data facts types, such as trends, outliers, and ranks, to form a data story. 
Most generated data facts describe statistical results without narratives to reflect how the data facts affect the status of the storyworld. 
For instance, \autoref{fig:case} A is an example of a data story generated by Calliope with six data facts. Only the first data fact (i.e., the total death over dates is decreasing) aligns with the conceptualization of a narrative event as it describes the changed status of the total death number. 
Furthermore, although some data storytelling conceptualizations pointed out that data storytelling involves not only data, none of the research in our corpus discussed how data findings can be combined with contextual information to form narrative events.

\subsection{Character}

\textit{Narratological perspectives.} When it comes to the status of characters in narrative, there are two general types of opinions. One is that the narrative is plot-centered, and the characters are secondary~\cite{chatman1978story}. Characters serve a functional role that connects the thread of a story~\cite{chatman1978story}. Characters are the agents around whom events happen, driving the narrative forward. The other is that narratives can also be character-centered, where the purpose of the narrative is revealing characterizations~\cite{todorov1977poetic,mckee2005story}. 
Whether a character is primary or secondary in a narrative, having human or human-like characters in a narrative is crucial for its affective appeal. Characters let the audience share their interests, antagonism, and emotions~\cite {chatman1978story,bal2009narratology}. 
Narrative theorists discuss characters in depth. For example, their classifications of characters as protagonists, antagonists, flat characters, and round characters provide a framework for understanding their various roles within stories and the nuanced emotional responses they evoke in the audience.

\textit{Data storytelling perspectives.}
However, it appears that the majority of current research on data storytelling does not emphasize characters. None of the data storytelling conceptualizations we coded mentioned characters. In our corpus, \textbf{we observed three types of understanding and processing of characters in data storytelling}. 

\textbf{The functional roles of characters can be taken by abstract objects or concepts.} As data findings often do not involve human or human-like agents, some data storytelling research indicates that abstract objects or concepts in data stories can also serve the role of traditional narrative characters in helping the audience connect threads.
Data characters are formally defined in recent research by Dasu et al.~\cite{a472dasu2023character} as \emph{visual entities that relate to the theme of the story and advance the story plot.} For example, in the graphics \textit{Iraq's bloody toll} that represents the death data in the United States military engagement in Iraq by an upside-down, red bar chart. Dasu coded the main character in this graphic as the red bars, as they represent the topic (i.e., death) throughout the entire story, as shown in \autoref{fig:case} B. 
Under this interpretation, to evoke affective responses, the design of data stories without human(-like) agents should leverage other storytelling techniques, such as sounds and narrative structure~\cite{lan2023affective}.
Note that Dasu et al.'s theory does not eliminate the situation in which a human or human-like agent is the character of a data story. They found that visual representations of main characters in data stories range from abstract presentations to images of real persons. This aligns with our following observations, in which human perspectives are revealed by characters in data stories.

\textbf{Add characterizations into the data story to achieve a similar effect to that of characters.} This approach to using characters in data storytelling includes human or human-like subjects, yet they do not act as agents who cause events to happen and carry the plot forward. Its manifestations may involve having a narrator appear on camera, incorporating images or illustrations of people, or describing individuals who are affected by the data. For instance, in the data video \textit{200 Countries, 200 Years, 4 Minutes versus 200 Years That Changed the World}, Hans Rosling, as the narrator, appears on camera to guide the narration and direct the audience's attention, as shown in \autoref{fig:case} C. 
This approach can achieve a similar effect to that of characters in traditional narratives by eliciting the audience's natural empathy towards entities that exhibit characterizations. In other words, it can leverage the inherent human tendency to connect with personified elements~\cite{bal2009narratology, mckee2005story}. 
A user study by Bradbury and Guadagno further supports this effect. They compared data videos with and without narrators on camera and found that most participants preferred to have a narrator as the character of the story~\cite{a68bradbury2020documentary}. 
However, the addition of characterizations to evoke emotional connections with the audience should be approached with caution. Previous studies have examined the effectiveness of anthropographics, which uses human-like icons to encode data, thereby giving the audience the impression that there appears to be a human presence behind the data. However, most of them struggle to find a significant effect of anthropographics on enhancing empathy~\cite{boy2017showing}. 
That's probably because data visualization here acts as a visual medium that leads the audience to the ``human'' dimension. However, there is no narrative that integrates characters and events to create a compelling story arc that can fully engage the audience's emotions and sense of empathy.

\textbf{A data story can center around a human(-like) protagonist.} This approach conveys data by telling how data shapes the living environments and affects the fates and decisions of the characters in the story.
For instance, Mittenentzwei et al.~\cite{a476mittenentzwei2023disease} explored whether the comprehension of disease data could be enhanced by integrating it into a story where an imaginary protagonist---a physician or patient---engages in defeating the disease, framed by the Hero's Journey story structure as shown in \autoref{fig:case} D.
Examples of narratives for revealing data are \emph{Emma, like 10\% of female CSVD patients, is between 50 and 60 years old...} and \emph{Ms. Schreiber observed that 10\% of her female patients and 20\% of her male patients...}
Another case is the data story \textit{Brooke Leave Home} by Concannon et al.~\cite{a39concannon2020brooke} for facilitating the understanding of data about the support available to young adults after they leave the care system in England. The story illustrates the challenges faced by and the support available to an imaginary protagonist named Brooke after she leaves the care system at age 18. Data insights are embedded into the narratives and visual scenes. For instance, the story mentions the number of care leavers when introducing Brooke's background. As Brooke makes purchases for her new home, the costs of major items and the depletion of Brooke's account are superimposed on the scene to comment on whether Brooke receives sufficient support, as shown in \autoref{fig:case} E.
In the user studies of both works, participants showed emotional responses to characters in the story.
While this approach fully leverages the functional and emotional roles of characters, data are still naturally more distant from many people's lives. The authors of \textit{Brooke Leave Home} applied the personalization technique by presenting data according to the audience's locations, which may be a valuable practice for future character-based data stories.

\subsection{Plot}
\textit{Narratological perspectives.} Aristotle defined a plot as the \emph{arrangement of incidents}~\cite{upton1748critical}. 
Chatman interpreted that a plot considers not only the sequence of events but also which events to emphasize and de-emphasize to focus on certain aspects of events and characters~\cite{chatman1978story}.
The sequence of events is not a simple, linear structure; instead, the events are connected; as Robert MacKee stated, \emph{plot is an accurate term that names the internally consistent, interrelated pattern of events that move through time to shape and design a story.}~\cite{mckee2005story} 
A well-designed plot not only narrates events but also generates dramatic effects such as suspense, surprise, conflicts, and resolutions~\cite{chatman1978story, mckee2005story, abbott2020cambridge}. 
Throughout the history of narrative, typical plots and their structures have been recognized as classic narrative structures for inspiring story creation, such as Freytag's Pyramid~\cite{freytag1895technique} and the Hero's Journey~\cite{campbell2008hero}.

\textit{Data storytelling perspectives.}
As discussed in Section~\ref{sec:technique}, not all data storytelling conceptualizations consider narrative structure to be an essential element for data storytelling. Overall, \textbf{we observed three types of understanding of plots in data storytelling}.

\textbf{Plots in data storytelling are not necessary.} 
This understanding of data storytelling aligns with the minimal conceptualization of narratives---the representation of events.
In such instances of data stories, data visualizations are often accompanied by narratives to reveal key data findings. However, the organization and interconnections of the data insights to form a narrative arc are often lacking.
For example, in the work~\cite{a38martinez2020data} that applies data storytelling to facilitate learning analytics, the authors give an example in which the key data insights are annotated in a timeline visualization of students' learning activities. These data insights, however, are not further organized and connected to draw any conclusion.

\textbf{Plots in data storytelling are the sequence of data facts connected by data relations.}
Starting from the work of Hullman et al.~\cite{hullman2011visualization} that classified the relationships (e.g., temporal and granularity) between adjacent visualizations in slideshow-style data stories, (semi-)automatic tools have been introduced to utilize such data relations to structure a data story~\cite{a12shi2020calliope,a24sun2022erato}.
These tools apply quantitative metrics like minimizing transition costs to evaluate the generated data stories.
As presented in \autoref{fig:case} A, in a data story generated by the tool Calliope~\cite{a12shi2020calliope}, adjacent data facts usually pertain to the same dimension or measures within a data table. This approach to structuring data stories can achieve certain logical coherence. However, as Lee et al.~\cite{lee2015more} discussed in their conceptualization, the organization of data story pieces should be geared towards higher-level communication goals like educating, entertaining, and persuading. Solely considering the statistical relationships for organizing data findings is insufficient for creating data stories with compelling structures.

\textbf{Plots in data storytelling should learn from classical narrative structures.} 
Two approaches to adapting classical narrative structures are observed.
This first is adapting plots from classical narrative structures that resonate with audience, such as the plot of the hero's growth and transformation after a series of challenges or temptations in the Hero's Journey structure~\cite{campbell2008hero}. 
For instance, the data story by Mittenentzwei et al.~\cite{a476mittenentzwei2023disease} in \autoref{fig:case} D is based on the premise of an imaginary protagonist to incorporate the Hero's Journey structure.
The second is drawing on the emotional journey provided by classical narratives, such as gradually intensified emotions in the rising action and climax in Freytag's Pyramid.
For example, in the design space for applying Freytag's Pyramid to data stories by Yang et al.~\cite{a17yang2021design}, the authors explored how the organization of data facts can elicit emotional responses, such as presenting contrasting data facts to create a twist in the plot for surprise.

\textbf{Summary.} Overall, our comparison reveals that data storytelling has made notable efforts to integrate traditional narrative theories. However, this integration is not seamless. Most fundamentally, data stories diverge from classical narratives at the level of the basic story elements. 
Moreover, as data storytelling increasingly develops alongside interactive techniques, narrative structures no longer necessarily depend on carefully arranged plot progressions to produce different experiences, in contrast to traditional narratives. 

\section{Discussion}
In this section, we discuss the implications for future research, followed by the limitations and future work of this study.

\subsection{Where Are We so Far?} 
Looking back on the advancement of data stories, we have witnessed the increasing number of research with various types of contributions and the flourishing of data stories in different application domains (Section \ref{sec:corpus_overview}).
A variety of goals and techniques are discussed in the conceptualizations of data storytelling (Section \ref{sec:conceptualization}).
Forms of data storytelling have expanded from presentations in 2D screens to embodied experiences in 3D environments, incorporating different sensory modalities, and the audience experiences have transformed from passive consumption to active participation (Section \ref{sec:forms}). 
In addition, data stories have established a close integration with narratology, where research has successfully borrowed established knowledge, such as narrative structures and character development, to improve the expressiveness of data stories (Section \ref{sec:concepts}). 
Some researchers have creatively applied narrative theories, for instance, by expanding traditional conceptions of character to develop theoretical frameworks for data as the narrative focus~\cite{a472dasu2023character}; while others are repurposing structural plot frameworks from fiction and drama for the physical exhibition of data, thereby guiding the design and refinement of visitor flow and experiential engagement~\cite{van2024situating}.
Together, these advancements show that data storytelling is a vibrant and evolving field. Through the continual integrating, adapting, and synthesizing of novel narrative concepts and techniques, it demonstrates itself to be a dynamic and self-renewing interdisciplinary field.

\subsection{Embrace the Diversity in Data Storytelling}
Parallel to the breakthroughs in data storytelling are the diverse conceptualizations. As suggested by our analysis, some conceptualizations even contradict each other in describing the essence of data storytelling, which can pose challenges for the community.
For example, some conceptualizations tightly bind data stories to highly specific but narrow techniques (e.g., rhetorical techniques, animations, annotations), which can obscure the broader possibilities and flexibility of storytelling approaches.
Also, the characterization of the target audience remains vague in much of the literature. Key terms like \emph{general} or \emph{non-technical audience} are often used loosely and without a clear definition, while most studies fail to account for distinctions in demographics, expertise, or context. This ambiguity makes it unclear for whom data stories are designed and in what scenarios they are meant to be effective.
Moreover, many conceptualizations embed diverse yet vaguely defined objectives such as \emph{entertain} or \emph{inspire action} without providing the means to measure them. This lack of clarity not only makes it difficult to align goals across different data stories, but, more fundamentally, it reflects the absence of a shared evaluative consensus within the field on what constitutes a ``good'' data story.
 
However, conceptual ambiguity can also signal opportunity. A highly relevant and illustrative precedent is the debate that emerged between traditional narratology and the new field of digital narrative powered by computational technology.
For instance, the rise of interactive fiction in the 1970s and hypertext fiction in the 1990s challenged conventional literary structures, while interactive films reimagined cinematic storytelling.
The advent of digital games further intensified the debate: narratologists sought to frame games within narrative theory, whereas ludologists argued for the primacy of interactivity and rule-based systems~\cite{koenitz2015interactive}. These long-standing discussions reflect not only highly diverse understandings of what constitutes \emph{narrative}, but also the generative, innovative potential of such debates.
For example, scholars such as Ryan \cite{ryan2006avatar} sought to reconcile different perspectives. She proposed that narrative should be understood not as a specific form, but as a cognitive storyworld model that can be evoked across media. From this perspective, narrativity becomes a matter of degree: any object capable of triggering such a mental construction possesses it to some extent.

To some extent, it was these bold explorations and theoretical controversies that stimulated innovation. By comparison, research on data storytelling remains at an early stage of theoretical development. Therefore, this research does \textbf{not} seek to establish a universal definition of data storytelling. Rather than prematurely enforcing conceptual closure, embracing plurality and productive disagreement may offer a more fertile path for the field’s long-term growth.

%

\subsection{Future Research Opportunities}

\textbf{Conversations Beyond Classic Narrative Theories}
While researchers in the visualization community have produced much meaningful and interesting work in dialogue with narrative concepts and theories, the conversation largely remains within the realm of ancient philosophy and traditional narratology. This encompasses theories from Freytag's pyramid structure~\cite{yang2021design} to Campbell's hero's journey~\cite{wei2024telling} and frameworks of theorists like Genette~\cite{lan2021understanding} and Chatman~\cite{Schroeder2023Telling}.
However, as has been widely acknowledged within broader narrative studies, traditional narrative theory often struggles to account for new narrative forms emerging under computational influence~\cite{koenitz2015interactive}. 
The various qualities introduced by technology, such as interactivity, spatiality, and co-authorship, all challenge classical models.

Future work could therefore consider engaging with narrative theories developed for or adapted to the digital age. For instance, Brenda Laurel~\cite{laurel2013computers}, building upon Aristotelian dramatic theory, proposed the Computer as Theatre theory. This theory frames interaction design as a form of theatrical performance, emphasizing the design of causality to guide the audience. In the context of data storytelling, this implies that designers must ensure that every user action during data exploration yields feedback that is logically coherent and emotionally intentional, making users feel they are participating in a meaningful, unified action. The Flying Wedge model she detailed as an example (where all possibilities are open at the start, choices gradually narrow the scope toward inevitability, culminating in a sense of cathartic completion) presents an intriguing narrative arc that is the exact inverse of the well-known Martini Glass model (which narrows before opening up) in data storytelling~\cite{a1segel2010narrative}. 
Similarly, Hanney~\cite{hanney2024one} critiqued the hero's journey structure, arguing it promotes a romanticized, individualistic heroism that is inherently patriarchal. Hanney then discussed several alternative frameworks, such as the Collective Journey and Queer Temporality, which could inform radically different designs and generative logics for data stories.

\noindent \textbf{Beyond Genres and Forms}
To address the issues of ambiguity, overlap, and limited flexibility inherent in genre and form classifications, we have intentionally proposed additional spectra to help break away from the constraints of typology-based thinking. Our spectra foreground properties that data story creators can actively tune: how it is situated in space and time (\textit{Spatiality}), how experience is ordered (\textit{Ordering}), how many \textit{Sensory modalities} are used to convey and interpret data, and how the audience participate in constructing meaning (\textit{Audience Co-authorship}). Because each spectrum can be adjusted independently, the framework supports \textit{combinatorial} thinking: data story creators can recombine choices across dimensions to imagine experiences onto an existing genre or new genre, and describe them precisely.

These spectra are not evaluative scales. 
Seen as a design blueprint, the four spectra help articulate recipes for different contexts. To give speculative examples, a museum installation might use a fixed spatial route. It could start with a short, repeatable linear introduction. It could then offer localized branching at stations for visitors who linger or return. An immersive AR experience might map story progression to locomotion. This makes ordering spatial, while keeping content modular. It can also layer audio cues or animated transitions to keep the narrative coherent during exploration. Interactive data comics suggest another combination. They often keep a readable panel sequence as a stable backbone. They then introduce nonlinearity in a controlled way. For instance, authors can append, replace, or jump between panels to support drill-down and alternative perspectives. They can also invite the audience to input. The audience might enter their own data, or talk with AI agent-powered characters to shape what appears next. Across such examples, the spectra encourage data story creators to treat ordering, spatiality, participation, and sensory richness as explicit, composable variables, and to justify where a design sits on each spectrum based on purpose, audience, and setting rather than adherence to any single genre.

\noindent \textbf{Rethink Opportunities of Human-AI Co-creation} 
\label{sec:co-creation}
Recent surveys have discussed research on tools for Human–AI co-creating data stories and proposed future opportunities \cite{li2024we, he2024leveraging, chen2023does}. They mainly take a data-storyteller-centered perspective, describing how AI can play roles such as creator, assistant, or reviewer \cite{li2024we} to support common phases like analysis, narration, visualization, and interaction \cite{he2024leveraging}, with efficiency and expressive generation as primary goals. Comparing what existing tools can support with the diversity of data stories revealed by our survey, we find that it is under-explored how AI may fundamentally change the forms of data stories rather than only improving existing workflows. 
For instance, AI enables chatbot-based stories and real-time generated stories, where the narrative can emerge through ongoing interaction instead of following a fully prewritten script—an idea aligned with discussions of AI-driven emergent narrative in interactive digital narrative research~\cite{koenitz2015interactive}. 
Furthermore, as discussed in Section \ref{sec:co-authorship}, the audience can become a co-author, such as in participatory \textbf{Events} and exploratory \textbf{Systems}. This introduces a third author (the audience) besides AI and the initial storyteller: AI may mediate participation, summarize inputs, or prioritize which contributions shape what happens next.
Finally, reflecting on Murray's \cite{murray1997hamlet} view of the future of narrative in cyberspace, which describes four core affordances---procedural, participatory, spatial, and encyclopaedic---of digital environments to shape a narrative. We have observed that the first three properties already map onto many data stories. The encyclopaedic potential using AI to organize large information spaces and let audiences drill down into rich detail while providing them with real-time data stories to understand data is a promising scenario.

\subsection{Limitations and Future Work}
Our goal is to provide a systematic and reproducible account of how research conceptualizes data storytelling within a well-defined publication scope. Accordingly, our corpus is derived from major scholarly indexing services and peer-reviewed venues, and should be interpreted as representative of work that is both discoverable through these infrastructures and explicitly framed using data storytelling-related terminology. This choice supports transparency and comparability, but it may under-represent perspectives that circulate through domain-specific outlets, non-English venues, or practice-oriented formats that are less consistently indexed. To mitigate this risk, we used multiple databases, de-duplicated results, and applied iterative coding that allowed new themes to emerge rather than forcing studies into a priori categories. We therefore position our findings as characterizing patterns observed in the curated corpus, while treating prevalence claims as corpus-dependent rather than field-exhaustive.

Future work can expand beyond English-language and highly indexed venues by incorporating regional publication outlets and practice-facing literature (for example, journalism, policy, and design communities), enabling comparisons of how conceptualizations vary across communities and sociotechnical contexts. Future work can also complement this literature-based synthesis with empirical methods, such as interviews, focus groups, and questionnaires, to examine how researchers across domains articulate, negotiate, and operationalize their conceptualizations of data visualization and data storytelling in practice.

\section{Conclusion}
In this paper, we synthesize how research conceptualizes data storytelling by surveying \npublication~publications from January 2010 to November 2025 and analyze the core questions of what data storytelling is, how it is constructed, why it is used, and who the audience is. We consolidate recurring conceptualizations into five perspectives, and we map the design space through a set of data story forms that capture the diversity of media, implementation styles, and research objectives. To move beyond form-based descriptions, we introduce four spectrum-based dimensions, ordering, sensory modalities, spatiality, and audience co-authorship, as analytic lenses for comparing experiences and reasoning about design trade-offs across contexts. Finally, by relating data storytelling to narratology through concepts such as event, character, and plot, we provide a complementary narrative framing that helps articulate what is distinctive about data-driven stories. Together, these contributions establish a shared vocabulary and a structured framework for describing, comparing, and designing data storytelling experiences, and they motivate future work that broadens coverage across communities and strengthens the empirical grounding of how conceptualizations are formed and operationalized in practice.


\bibliographystyle{eg-alpha-doi} 
\bibliography{main}  

\newpage
\section{Short Biographies of Authors}
\textbf{Leni Yang} is a post-doc research fellow in Bivwac, Inria, CNRS. Her research interests include Data Visualization and Human-Computer Interaction with a focus on investigating effective design theories~\cite{yang2023understanding,yang2021design}, design strategies~\cite{zhu2024reader}, and authoring tools~\cite{wang2024outlinespark,lin2023inksight,takahira2025tangiblenet} for turning complex data into understandable and compelling data stories.\\

\noindent \textbf{Zezhong Wang}, is a postdoctoral researcher at the Interactive Experiences Lab, Simon Fraser University, Canada. His research interests encompass the creation and examination of data-driven storytelling, harnessing visual communication and insights derived from data, and employing design innovation to enhance data accessibility and engagement for the general public. Zezhong earned his PhD from the School of Informatics, University of Edinburgh, where he explored the creation of data comics. His research delved into the effectiveness of data comics~\cite{wang2019comparing}, their creation~\cite{wang2019teaching, wang2021reporting, wang2022interactive}, and materials for teaching visualization literacy~\cite{wang2020cheat}, earning him the VGTC Visualization Dissertation Award Honorable Mention in 2023. \\

\noindent \textbf{Sheelagh Carpendale} is a professor at Simon Fraser University, Canada, where she holds a Canada Research Chair in Data Visualization and is a Fellow of the Royal Society of Canada. Her many awards include the IEEE Visualization Career Award, CS-CAN Lifetime Achievement, a BAFTA, and being inducted into both the ACM CHI (Computer-Human-Interaction) Academy and the IEEE Visualization Academy. As an international leader in information visualization and interaction design, she studies how people interact with information both in work and social settings. She works towards designing more natural, accessible, and understandable interactive visual representations of data.

\noindent \textbf{Xingyu Lan} is an assistant professor in the School of Journalism, Fudan University, Shanghai, China. Her research focuses on data storytelling, audience experience, and human-AI interaction. She conducted a series of studies about the design spaces and affective design of data stories~\cite{lan2021kineticharts,lan2021smile,lan2022negative,lan2021understanding} and received awards such as IEEE VIS Best Paper Award~\cite{lan2023affective} and Honorable Mention Award~\cite{lan2024came}. She was also an award-winning practitioner and has rich experience in creating data stories.\\

\newpage


\end{document}